\documentstyle[preprint,epsfig,aps]{revtex}
\begin{document}
\draft

\preprint{\begin{tabular}{l}
\hbox to\hsize{\hfill {\bf (hep-ph/9905528)}}\\[-3mm]
\hbox to\hsize{\hfill YUMS 99-011}\\[-3mm]
\hbox to\hsize{\hfill HUPD 9905}\\[-3mm]
\vspace{1.0cm}
\end{tabular} }
%
\title{New Physics Effects in $ B \rightarrow K^{(*)} \nu {\nu}$ Decays}
\author{C. S. Kim$^1$\thanks{kim@kimcs.yonsei.ac.kr,~~http://phya.yonsei.ac.kr/{}~cskim/},
Yeong Gyun Kim$^1$\thanks{ygkim@cskim.yonsei.ac.kr}
and T. Morozumi$^2$\thanks{morozumi@theo.phys.sci.hiroshima-u.ac.jp}}
\address{1.$~$Dept. of Physics, Yonsei University, Seoul 120-749, Korea}
\address{2.$~$Dept. of Physics, Hiroshima University,
739-8526 Higashi-Hiroshima, Japan}
\maketitle
\def\beq{\begin{equation}}
\def\eeq{\end{equation}}
\def\bea{\begin{eqnarray}}
\def\eea{\end{eqnarray}}
\def\nn{\nonumber}
\def\bpp{{\bf p'}}
\def\bk{B \rightarrow K}
\def\bks{B \rightarrow K^*}
\def\nub{{\bar \nu}}
\def\nunu{\nu \nu}
\def\nunub{\nu \nub}
\def\nubnub{\nub \nub}
\def\bknunub{\bk \nunub}
\def\bksnunub{\bks  \nunub}
\def\bknunu{\bk \nunu (\nubnub)}
\def\bksnunu{\bks \nunu (\nubnub)}
\vspace{0.5cm}
\begin{center}
 (\today)
\end{center}

\begin{abstract}
\noindent 
We present a model-independent analysis of exclusive rare $B$ decays, $B
\rightarrow K^{(*)} \nu \nu$.   The effect of possible new physics is
written in terms of dimension-$6$ four-fermi interactions. 
The lepton number violating scalar- and tensor-type interactions 
are included, and they induce
$ B \rightarrow K^{(*)} \nu \nu ({\bar \nu} {\bar \nu}) $ decays.
We show systematically how the branching ratios and 
missing mass-squared spectrum
depend on the coefficients of the four-fermi interactions.
\end{abstract}

\newpage
\baselineskip .29in

\section{Introduction}

\noindent
Flavor-changing-neutral-current (FCNC) process $b \rightarrow s \nu {\bar \nu}$ 
is a theoretically very clean mode in the Standard Model (SM) \cite{buras}. 
However, it might be  extremely difficult to measure precisely
the inclusive mode $B \rightarrow X_s \nu \bar \nu$ because
it requires to reconstruct all $X_s$ (together with two neutrinos). 
Experimentally it could be much easier to
measure the exclusive modes $B \rightarrow K^{(*)} \nu {\bar \nu}$.
The corresponding processes in $K$-meson system are 
$K_L \rightarrow \pi^0 \nu \bar \nu$ and $K^+ \rightarrow  \pi^+ \nu \bar \nu$,
and the expected branching ratios are $10^{-10}$ to $10^{-11}$ \cite{buras,nir}.
Compared with the rare decays of $K$-meson,
the branching fractions of the $B$-meson decays are much larger, and
the theoretical estimates are  $\sim 10^{-5}$ for
$B \rightarrow K^{*} \nu \bar{\nu}$ case and $\sim 10^{-6}$ 
for $B \rightarrow K \nu \bar{\nu}$ case \cite{grossman,aliev}.
The form factors of the decay process $K \rightarrow \pi \nu \bar{\nu}$ in the SM
are related to the well known $K_{l3}$ decay with isospin symmetry \cite{buras}.
In the $B$ system, while the form factors of $B \rightarrow \rho(\pi) \nu \bar{\nu}$
are directly related to those of $B \rightarrow \rho(\pi) l \nu$ decay, 
for $B \rightarrow K^{(*)}  \nu \bar{\nu}$ processes they are only 
related to those of $B \rightarrow \rho(\pi) l \nu$ in the SU(3) limit \cite{aliev}.
Therefore, we still have to rely on models to estimate the form factors.
Although this may introduce some model dependence of the hadronic form factors,   
it is still worth studying the exclusive decays, 
$B \rightarrow K^{(*)} \nu {\bar \nu}$.
Because of the higher statistics, 
we could study not only the branching fraction but also
the distributions, like missing mass-squared spectrum.

In this paper, we investigate the possible new physics effects on
the branching ratio and the spectrum of $B \rightarrow K^{(*)} \nu \nu$ decays.
The spectrum is sensitive to the types of the interactions and
is useful for discriminating the various new physics effects \cite{meli1}.
We assume that the new physics effects are parametrized by
dimension-$6$ four-fermi interactions. (See Ref. \cite{CKIM} for the most general
analysis of inclusive decays $ B \rightarrow X_s l^+ l^-$.)
Further, we assume that
the right-handed component of (anti-)neutrinos is supplied by charge
conjugated field of the left-handed neutrinos.
In the SM, only one operator
with the structure $(V-A)_{\it quarks} \times (V-A)_{\it neutrinos}$
contributes to the process. In extension of the SM, but still within
chirality conserving four-fermi interaction,  another structure
$(V+A)_{\it quarks} \times (V-A)_{\it neutrinos}$ is possible. 
Further, including the chirality changing interactions,
the lepton number violating operators with the types 
$S_{\it quarks} \times S_{\it neutrinos}$ ($S$= scalar-type interactions) and
$T_{\it quarks} \times T_{\it neutrinos}$ ($T$= tensor-type interactions)
are also possible.
The scalar and vector interactions were studied
in the context of $K \rightarrow  \pi \nunub$ in a left-right model in  Ref. \cite{KIYO}.

Thus the most general model independent Lagrangian is given by,
\begin{eqnarray}
{\cal L} = \frac{\sqrt{2} G_F \alpha}{\pi}
           &\{& ( C_{Lij}^V \overline{s_L} \gamma_\mu b_L +
           C_{Rij}^V \overline{s_R} \gamma_\mu b_R)
               ( \overline{\nu_{Li}} \gamma^\mu \nu_{Lj} ) \nonumber \\
           &+& ( C_{Rij}^S \overline{s_L} b_R + C_{Lij}^S \overline{s_R} b_L )
               ( \overline{(\nu_{Li})^C} \nu_{Lj} ) \nonumber \\
           &+& ( \tilde{C}_{Rij}^S \overline{s_L} b_R +
           \tilde{C}_{Lij}^S \overline{s_R} b_L )
               ( \overline{\nu_{Li}} (\nu_{Lj})^C ) \nonumber \\
           &+&   C_{Lij}^T (\overline{s_R} \sigma_{\mu \nu} b_L)
                       (\overline{(\nu_{Li})^C} \sigma^{\mu \nu} \nu_{Lj})
                       \nonumber \\
           &+&   C_{Rij}^T (\overline{s_L} \sigma_{\mu \nu} b_R)
                       (\overline{\nu_{Li}} \sigma^{\mu \nu} (\nu_{Lj})^C)~ + h.~c.\},
\label{eq:1}
\end{eqnarray}
where the neutrino species are denoted by ${i,j}$.
{}From Eq.(\ref{eq:1}), we note the following points:\\
(i)  $C_L^V$, $C_R^V$ terms contribute to
$B \rightarrow K \nu \bar{\nu}$ process,\\
(ii)  $C_L^S$, $C_R^S$, $C_L^T$ terms contribute to
$B \rightarrow K \bar{\nu} \bar{\nu}$ process, and\\
(iii)  $\tilde{C}_L^S$, $\tilde{C}_R^S$,$C_R^T$ terms contribute to
$B \rightarrow K \nu \nu$ process.\\
In the Appendix, we derive the statistical factors for the production of
(even theoretically) indistinguishable final state
neutrinos, {\it i.e.},  $B \to K^{(*)} {\nu_i}{\nu_i}$ and
$B \to K^{(*)} \bar{\nu_i}\bar{\nu_i}$.

\section{Form Factors}

\noindent
First we write the form factors for $\bk$ case
as follows,
\begin{eqnarray}
<K(p')| \bar{s} \gamma_\mu b | \overline{B} (p)> &=&
f_+ (p+p')_\mu + f_- (p-p')_\mu, \label{eq:2} \\
<K(p')| \bar{s} \sigma_{\mu \nu} b | \overline{B} (p)> &=&
i \frac{f_T}{m_B} [(p+p')_\mu (p-p')_\nu - (p-p')_\mu (p+p')_\nu].
\label{eq:3}
\end{eqnarray}
{}From Eq. (\ref{eq:2})  the scalar form factor is obtained,
\begin{eqnarray}
<K(p')| \bar{s} b | \overline{B} (p)>
&=& \frac{1}{m_b-m_s} [ f_+ (m_B^2-m_K^2) + f_- q^2 ].
\label{eq:4}
\end{eqnarray}
In the rest frame of $B$-meson, $p=(m_B,0)$ and $p'=(\sqrt{{\bf p'}^2+m_K^2},
{\bf p'})$, and
\begin{eqnarray}
p+p'=(m_B+\sqrt{{\bpp}^2+m_K^2}, {\bpp}) ,\quad
q=p-p'=(m_B-\sqrt{{\bpp}^2+m_K^2},-{\bpp}).
\label{eq:5}
\end{eqnarray}
The matrix element of the tensor operator in the  $B$-meson rest frame is
given by
\begin{eqnarray}
<K(p')| \bar{s} \sigma^{0 i} b | \overline{B} (p)>
= -2i {p'}^i f_T ,
\label{eq:6}
\end{eqnarray}
where all the other components are zero. Therefore, near the zero-recoil
the tensor form factor is suppressed by a factor of $(p'/m_B)$ compared with that of
the scalar operator.
The form factors for $\bks$ are written in the same way,
\begin{eqnarray}
<K^*(p',\epsilon)| \bar{s} \gamma_\mu b | \overline{B} (p)> &=&
i g \epsilon_{\mu \nu \lambda \sigma}
\epsilon^{* \nu} (p+p')^\lambda (p-p')^\sigma,
\label{eq:7} \\
<K^*(p',\epsilon)| \bar{s} \gamma_\mu \gamma_5b | \overline{B} (p)> &=&
f \epsilon^*_\mu + a_+ (\epsilon^* \cdot p) (p+p')_\mu
                 + a_- (\epsilon^* \cdot p) (p-p')_\mu ,
\label{eq:8} \\
<K^*(p',\epsilon)| \bar{s} \sigma_{\mu \nu} b | \overline{B} (p)>
&=& g_+ \epsilon_{\mu \nu \lambda \sigma}
\epsilon^{* \lambda} (p+p')^\sigma +
g_- \epsilon_{\mu \nu \lambda \sigma}
\epsilon^{* \lambda} (p-p')^\sigma \nonumber \\
&+& h \epsilon_{\mu \nu \lambda \sigma} (p+p')^\lambda (p-p')^\sigma
(\epsilon^* \cdot p).
\label{eq:9}
\end{eqnarray}
{}From Eq. (\ref{eq:7})-(\ref{eq:8}),
we obtain the scalar and pseudoscalar form factors, respectively
\begin{eqnarray}
<K^*(p',\epsilon)| \bar{s} b | \overline{B} (p)>&=&0,
\label{eq:10} \\
<K^*(p',\epsilon)| \bar{s} \gamma_5 b | \overline{B} (p)>
&=& \frac{-1}{m_b+m_s}
[f (\epsilon^* \cdot p) + a_+ (\epsilon^* \cdot p) (m_B^2-m_{K^*}^2)
                        + a_- (\epsilon^* \cdot p) q^2 ].
\label{eq:11}
\end{eqnarray}

For numerical calculations of the $B \rightarrow K, K^*$ transition form factors, 
we use a dispersion quark model calculation \cite{MELI} in the whole
kinematic range of $q^2$ with the parametrization 
$$
f_i(q^2)=\frac{f_i(0)}{1-\sigma_1 q^2 + \sigma_2 q^4} . 
$$
In Ref. \cite{MELI}, the authors adopt
the quark masses and the wave functions of the Godfrey-Isgur (GI) model \cite{GI}
for the hadron spectrum with a switched-off one-gluon exchange (OGE) potential.
It is found that the resulting form factors are in good agreement with 
the lattice simulations at large $q^2$.
For convenience, we present the simple fit results of the GI-OGE model,
$f_i(0),~ \sigma_1$ and $\sigma_2$, in Table I.

\section{Missing mass-squared spectrum and  branching ratios }

\noindent
Here we assume the mass of neutrinos to be zero, and therefore, we
neglect possible effects of neutrino mass in the spectrum and the branching ratios.
Now let us derive the missing mass-squared spectrum  ${d\Gamma / dq^2}$.
In this Section we show the results for the different flavor cases, {\it i.e.},
$B \to K^{(*)} \nu^{i} {\bar \nu^{j}} (i \neq j)$,
$B \to K^{(*)} {\bar \nu^{i}} {\bar \nu^{j}} (i \neq j)$ and 
$B \to K^{(*)} {\nu^{i}} {\nu^{j}} (i \neq j)$.
For the cases $i=j$, we then get the results from:
\begin{itemize}
\item
For $B \to K^{(*)} \nu {\bar \nu}$, the results are the same as the case with $i \neq j$.
\item
For $B \to K^{(*)} \nu {\nu}$ and $B \to K^{(*)} {\bar \nu} {\bar \nu}$,
the results should be multiplied by the statistical factor two.
\end{itemize}
The derivation of the statistical factors for the production of
(theoretically) indistiguishable final state neutrinos
is given in the Appendix. 

We first study  $\bk$ case, and the missing mass-squared spectra are given as,
\begin{eqnarray}
\frac{d\Gamma(B \rightarrow K \nu \bar{\nu})}{d q^2} =
\frac{G_F^2 \alpha^2 }{96 \pi^5} |C_L^V + C_R^V|^2~ f_+^2 |p'|^3,
\label{eq:12}
\end{eqnarray}
\begin{eqnarray}
\frac{d\Gamma(B \rightarrow K \bar{\nu} \bar{\nu})}{d q^2}
&=&
\frac{G_F^2 \alpha^2 }{256 \pi^5} |C_L^S + C_R^S|^2
\frac{|p'| q^2}{m_B^2 (m_b-m_s)^2} (f_+ (m_B^2-m_K^2)+f_- q^2)^2
\nonumber \\
&+&
\frac{G_F^2 \alpha^2}{48 \pi^5} |C_L^T|^2
\frac{f_T^2}{m_B^2} |p'|^3 q^2,
\label{eq:13}
\end{eqnarray}
\begin{eqnarray}
\frac{d\Gamma(B \rightarrow K \nu \nu)}{d q^2}
&=&
\frac{G_F^2 \alpha^2 }{256 \pi^5} |\tilde {C}_L^S + \tilde{C}_R^S|^2
\frac{|p'| q^2}{m_B^2 (m_b-m_s)^2} (f_+ (m_B^2-m_K^2)+f_- q^2)^2
\nonumber \\
&+&
\frac{G_F^2 \alpha^2}{48 \pi^5} |C_R^T|^2
\frac{f_T^2}{m_B^2} |p'|^3 q^2,
\label{eq:14}
\end{eqnarray}
where
$|\bpp|$ is the three-momentum magnitude of $K$ in the $B$-meson rest frame
and can be written as,
\begin{equation}
|{\bpp}|=\frac {\sqrt{\lambda({m_B}^2,{m_K}^2,q^2)}}{2 m_B},
\label{eq:15}
\end{equation}
where $\lambda(a,b,c)=a^2+b^2+c^2-2 a b -2 b c-2 c a$.
The flavor indices in $C$'s are suppressed and should be read as: 
$C^V_{Lij}=C^V_L (i\ne j) ,C^V_{Rij}=C^V_R (i\ne j),
C_{Lij}^T=C_L^T (i\ne j), C_{Rij}^T=C_R^T (i \ne j),
\tilde{C}_{Lij}^T=\tilde{C}_L^T (i\ne j),  \tilde{C}_{Rij}^T=\tilde{C}_R^T (i \ne j).$
Summing all three contributions, Eqs. (12)-(14),
the total differential decay rate is given by,
\begin{eqnarray}
\frac{d\Gamma(B \rightarrow K )}{d q^2} =
&|& C_L^V + C_R^V|^2~ V_K (q^2) + \nonumber \\
&(& |C_L^S + C_R^S|^2 + |\tilde {C}_L^S + \tilde{C}_R^S|^2 )~S_K (q^2) +
( |C_L^T|^2 + |C_R^T|^2 )~ T_K (q^2),
\label{eq:16}
\end{eqnarray}
where
\begin{eqnarray}
V_K (q^2) &=&
\frac{G_F^2 \alpha^2 }{96 \pi^5}~ f_+^2 |p'|^3,
\label{eq:17}
\\
S_K (q^2) &=&
\frac{G_F^2 \alpha^2 }{256 \pi^5}
\frac{|p'| q^2}{m_B^2 (m_b-m_s)^2} (f_+ (m_B^2-m_K^2)+f_- q^2)^2,
\label{eq:18}
\\
T_K (q^2) &=&
\frac{G_F^2 \alpha^2}{48 \pi^5}
\frac{f_T^2}
{m_B^2} |p'|^3 q^2.
\label{eq:19}
\end{eqnarray}

The end points of the phase space, {\it i.e.}, $q^2=0$ and $q^2=(m_B-m_K)^2$,
correspond to
$|{\bpp}| = (m_B^2-m_K^2)/(2~ m_B)$ (maximal-recoil)
and  $|{\bpp}| =0$ respectively (zero-recoil).
The characteristic dependence on the kinematical variables
$q^2$ and $|p'|$ in Eq. (\ref{eq:17})-(\ref{eq:19})
can be seen in Fig. 1(a).
$V_K (q^2)$,~$S_K (q^2)$ and $T_K (q^2)$ correspond to
the solid, dashed and dotted line, respectively.
Ignoring the momentum dependence of the form factors,
near the minimum of the missing mass-squared,
the spectrum due to tensor- and scalar-type interactions
linearly grows as $q^2$ increases while the spectrum of
vector-type interactions  approaches a non-zero constant as $q^2
\rightarrow 0$. This is related to the fact that the collinear neutrino and
anti-neutrino have zero total helicity while the collinear
(anti-)neutrino and (anti-)neutrino have
$-(+)1$ total helicity. At the maximum-recoil limit
of $K$-meson,
the conservation of the total helicity cannot be satisfied for the case of
two neutrinos or two anti-neutrinos in the final states.
Near the end point of the spectrum,
the first derivative of the spectrum due to the scalar interactions
becomes infinity while that of the other spectra becomes zero.
The sharp rise of the spectrum of the scalar interaction near the
zero-recoil of $K$-meson is related to the fact that the back-to-back 
(anti-)neutrino and (anti-)neutrino have zero total helicity.
The suppression of the spectrum occurs for the vector interactions
because the back-to-back anti-neutrino and neutrino have $\pm1$
helicities and the helicity conservation cannot be satisfied.
As for the spectrum of the tensor interaction near the zero-recoil,
there is a suppression factor of $|\bpp|^2$ compared with that
of the scalar interaction. 

We now turn to $B \rightarrow K^*$ case.
By setting $\epsilon=\epsilon_L=({|\bpp|},0,0,E_K)$ (longitudinal polarization),
or $\epsilon=\epsilon_{T}$ (transverse polarization),
we can show that the following matrix elements vanish,
\begin{eqnarray}
<K^*(p',\epsilon_L)| \bar{s} \gamma_\mu b | \overline{B} (p)>=0
, \quad
<K^*(p',\epsilon_{T})| \bar{s} \gamma_5 b | \overline{B} (p)>=0.
\label{eq:20}
\end{eqnarray}
First let us consider the case of longitudinally polarized $K^*$,
\begin{eqnarray}
\frac{d\Gamma_L (B \rightarrow K^* \nu \bar{\nu})}{d q^2}
= \frac{G_F^2 \alpha^2}{384 \pi^5} |C_L^V - C_R^V|^2
\frac{|p'|}{m_B^2 m_{K^*}^2} (f(m_B E'-m_{K^*}^2)+2 a_+ m_B^2 |p'|^2)^2,
\label{eq:21}
\end{eqnarray}
\begin{eqnarray}
\frac{d\Gamma_L (B \rightarrow K^* \bar{\nu} \bar{\nu})}{d q^2}
&=& \frac{G_F^2 \alpha^2}{256 \pi^5} |C_L^S - C_R^S|^2
\frac{|p'|^3 q^2}{m_{K^*}^2 (m_b+m_s)^2}
( f + a_+ (m_B^2-m_{K^*}^2) + a_- q^2 )^2
\nonumber \\
+\frac{G_F^2 \alpha^2}{48 \pi^5} |C_L^T|^2 &\times&
\frac{|p'| q^2}{m_B^2 m_{K^*}^2}
(g_+ (m_B E'+ m_{K^*}^2)+g_- (m_B E' - m_{K^*}^2)+2~ h~ m_B^2 |p'|^2 )^2,
\label{eq:22}
\end{eqnarray}
\begin{eqnarray}
\frac{d\Gamma_L (B \rightarrow K^* \nu \nu)}{d q^2}
&=& \frac{G_F^2 \alpha^2}{256 \pi^5} |\tilde{C}_L^S - \tilde{C}_R^S|^2
\frac{|p'|^3 q^2}{m_{K^*}^2 (m_b+m_s)^2}
( f + a_+ (m_B^2-m_{K^*}^2) + a_- q^2 )^2
\nonumber \\
+ \frac{G_F^2 \alpha^2}{48 \pi^5} |C_R^T|^2 &\times&
\frac{|p'| q^2}{m_B^2 m_{K^*}^2}
(g_+ (m_B E'+ m_{K^*}^2)+g_- (m_B E' - m_{K^*}^2)+2~ h~ m_B^2 |p'|^2 )^2.
\label{eq:23}
\end{eqnarray}

The total differential decay rate, Eqs. (21)-(23), is given by
\begin{eqnarray}
\frac{d\Gamma(B \rightarrow K^* )_L}{d q^2} &\equiv&
\frac{d\Gamma(B \rightarrow K^*_{h=0} )}{d q^2} =
| C_L^V - C_R^V|^2~ V_L (q^2) + \nonumber \\
&(& |C_L^S - C_R^S|^2 + |\tilde {C}_L^S - \tilde{C}_R^S|^2 )~S_L (q^2) +
( |C_L^T|^2 + |C_R^T|^2 )~ T_L (q^2),
\label{eq:24}
\end{eqnarray}
where
\begin{eqnarray}
V_L (q^2) &=&
\frac{G_F^2 \alpha^2}{384 \pi^5}
\frac{|p'|}{m_B^2 m_{K^*}^2} (f(m_B E'-m_{K^*}^2)+2 a_+ m_B^2 |p'|^2)^2,
\label{eq:25}
\\
S_L (q^2) &=&
\frac{G_F^2 \alpha^2}{256 \pi^5}
\frac{|p'|^3 q^2}{m_{K^*}^2 (m_b+m_s)^2}
( f + a_+ (m_B^2-m_{K^*}^2) + a_- q^2 )^2,
\label{eq:26}
\\
T_L (q^2) &=&
\frac{G_F^2 \alpha^2}{48 \pi^5}
\frac{|p'| q^2}{m_B^2 m_{K^*}^2}
(g_+ (m_B E'+ m_{K^*}^2)+g_- (m_B E' - m_{K^*}^2)+2~ h~ m_B^2 |p'|^2 )^2.
\label{eq:27}
\end{eqnarray}
In Fig. 1(b), we show $V_L (q^2)$, $S_L (q^2)$ and $T_L (q^2)$,
which correspond to
the solid, dashed and dotted line, respectively.
For the large-recoil limit, {\it i.e.}, $q^2 \rightarrow 0$, the spectrum
is similar to that of the $B \rightarrow K$ case. Near the zero-recoil 
point, the sharp rise of the spectrum for the vector- and
tensor-type interactions is observed while the spectrum of the scalar-type
interaction is suppressed.  

Now we turn to the case of transversely polarized $K^*$.
For this case, the vector- and tensor-type
interactions contribute to the process.
\begin{eqnarray}
\frac{d\Gamma_{(\pm)} (B \rightarrow K^* \nu \bar{\nu})}{d q^2}
= \frac{G_F^2 \alpha^2}{384 \pi^5} \frac{|p'| q^2}{m_B^2}
|~ (C_L^V+C_R^V) 2~g~m_B |p'| \mp (C_L^V-C_R^V)~f~ |^2,
\label{eq:28}
\end{eqnarray}
\begin{eqnarray}
\frac{d\Gamma_{(\pm)} (B \rightarrow K^* \bar{\nu} \bar{\nu})}{d q^2}
= \frac{G_F^2 \alpha^2}{48 \pi^5} |C_L^T|^2
\frac{|p'|}{m_B^2} (2~g_+~ m_B |p'| \pm (g_+ (m_B^2-m_{K^*}^2)+g_- q^2))^2,
\label{eq:29}
\end{eqnarray}
\begin{eqnarray}
\frac{d\Gamma_{(\pm)} (B \rightarrow K^* \nu \nu)}{d q^2}
= \frac{G_F^2 \alpha^2}{48 \pi^5} |C_R^T|^2
\frac{|p'|}{m_B^2} (2~g_+~ m_B |p'| \pm (g_+ (m_B^2-m_{K^*}^2)+g_- q^2))^2,
\label{eq:30}
\end{eqnarray}
where $|p'|$ and $E'$ are the $K^*$ three-momentum magnitude
and energy in the $B$-meson rest frame.

The total differential decay rate, Eqs. (28)-(30), is given by
\begin{eqnarray}
\frac{d\Gamma (B \rightarrow K^* )_+}{d q^2} &\equiv&
\frac{d\Gamma (B \rightarrow K^*_{h=+1})}{d q^2}
= |C_L^V|^2~ V_1 (q^2) + |C_R^V|^2~ V_2 (q^2)
+ Re(C_L^V C_R^{V*})~V_3 (q^2) \nonumber \\
&+& (|C_L^T|^2 + |C_R^T|^2)~ T_+ (q^2),
\label{eq:31}
\end{eqnarray}
\begin{eqnarray}
\frac{d\Gamma (B \rightarrow K^* )_-}{d q^2} &\equiv&
\frac{d\Gamma (B \rightarrow K^*_{h=-1})}{d q^2}
= |C_L^V|^2~ V_2 (q^2) + |C_R^V|^2~ V_1 (q^2)
+ Re(C_L^V C_R^{V*})~V_3 (q^2) \nonumber \\
&+& (|C_L^T|^2 + |C_R^T|^2)~ T_- (q^2),
\label{eq:32}
\end{eqnarray}
where
\begin{eqnarray}
V_1 (q^2) &=&
\frac{G_F^2 \alpha^2}{384 \pi^5} \frac{|p'| q^2}{m_B^2}
(2~g~m_B |p'| - f)^2,
\label{eq:33}
\\
V_2 (q^2) &=&
\frac{G_F^2 \alpha^2}{384 \pi^5} \frac{|p'| q^2}{m_B^2}
(2~g~m_B |p'| + f)^2,
\label{eq:34}
\\
V_3 (q^2) &=&
\frac{G_F^2 \alpha^2}{384 \pi^5} \frac{|p'| q^2}{m_B^2}
2 (4 g^2 m_B^2 |p'|^2 - f^2),
\label{eq:35}
\\
T_\pm (q^2) &=&
\frac{G_F^2 \alpha^2}{48 \pi^5}
\frac{|p'|}{m_B^2} (2~g_+~ m_B |p'| \pm (g_+ (m_B^2-m_{K^*}^2)+g_- q^2))^2.
\label{eq:36}
\end{eqnarray}
In Figs. 1(c) and (d), we show $V_1$, $V_2$, $V_3$, $T_-$ and $T_+$.
$V_1 (q^2)$, $V_2 (q^2)$, $V_3 (q^2)$ and $T_- (q^2)$ correspond to
the solid, dashed, dotted and dot-dashed line,
respectively, in Fig. 1(c), and $T_+ (q^2)$ corresponds to the solid line in Fig. 1(d).

Note that in real experiments we cannot be able to distinguish the transverse
polarization $h=+1$ from $h=-1$ due to the non-detection of the two neutrinos. 
Therefore, we have to add two transverse polarizations,
\begin{eqnarray}
\frac{d\Gamma (B \rightarrow K^*)_T}{d q^2}
\equiv \frac{d\Gamma (B \rightarrow K^*_{h=+1})}{d q^2}
+ \frac{d\Gamma (B \rightarrow K^*_{h=-1})}{d q^2} .
\label{eq:37}
\end{eqnarray}
We note that Eq. (\ref{eq:37}) is symmetric under the interchange of the
variables $C_L$ and $C_R$. Thus, we cannot distinguish the interactions
with the opposite chirality structure using the spectrum. 
This contrasts with Ref. \cite{meli1},
where the asymmetry between $B \to K^*_{h=+1}$ and $B \rightarrow K^*_{h=-1}$
is assumed to be experimentally observed, thus leading to 
the measurement of the $C_L$ and $C_R$ separately.

\section{Effect of new interactions}

\noindent
In order to show the sensitivity of the branching ratios to the
new physics effects, we show the dependence of the branching ratios
on each coefficient.  For the numerical computation of
the branching ratio, 
we assume that there are three flavors of neutrinos, $\nu_{e,\mu,\tau}$,
and the interactions in Eq. (1) are universal and diagonal on the neutrino flavors,
{\it i.e.}, $C_{L,Rij}=C_{L,R} \delta_{ij}$ and $\tilde{C}_{L,Rij}=\tilde{C}_{L,R} \delta_{ij}$.
Therefore, we multiply three for the $\nunub$ final states and multiply six
$(= 3 \times 2)$ for the $\nunu (\nubnub)$ final states. (See Appendix for the
statistical factors of theoretically  indistinguishable neutrinos.)

The dependence of the branching ratios ${\cal BR}(B \rightarrow K)$,
${\cal BR}(B \rightarrow K^*_{h=0})$, 
${\cal BR}(B \rightarrow K^*_{h=+1}) + {\cal BR}(B \rightarrow K^*_{h=-1})$ 
and ${\cal BR}(B \rightarrow K^*_{h=+1}) + {\cal BR}(B \rightarrow K^*_{h=-1}) 
+ {\cal BR}(B \rightarrow K^*_{h=0})$ 
on the coefficients $C_X$ are
shown in Fig. 2(a), 2(b), 2(c) and 2(d), respectively.
Here, the dependence on $C_X = \tilde{C}_L^V / C_L^V ({\rm SM})$, $C_R^V / C_L^V ({\rm SM})$,
$C_L^S / C_L^V ({\rm SM})$
and $C_L^T / C_L^V ({\rm SM})$ are corresponds to 
the solid, dashed, dotted
and dot-dashed line, respectively, where
$\tilde{C}_L^V \equiv C_L^V - C_L^V ({\rm SM})$.
In order to calculate the branching fraction of each process,
we use the averaged lifetime of $B^\pm$ and $B^0$  
from Particle Data Book~\cite{PDBook},
$$
\tau_{B^\pm} = (1.62\pm 0.06) \times 10^{-12} sec , ~~~{\rm and}~~~
\tau_{B^0} = (1.56\pm 0.06) \times 10^{-12} sec . 
$$

Fig. 3 shows the dependence of
the ratio $R$ of produced $K$ to $K^*_T$ mesons, defined in \cite{meli1} as
\begin{eqnarray}
R \equiv \frac{{\cal BR}(B \rightarrow K)}
{{\cal BR}(B \rightarrow K^*_{h=-1})+{\cal BR}(B \rightarrow K^*_{h=+1})}
\end{eqnarray}
on the $C_X$, respectively.
Here $C_X = \tilde{C}_L^V / C_L^V ({\rm SM})$, $C_R^V / C_L^V ({\rm SM})$,
$C_L^S / C_L^V ({\rm SM})$
and $C_L^T / C_L^V ({\rm SM})$, and the dependence on these coefficients corresponds to
the solid, dashed, dotted
and dot-dashed line, respectively. 

In Figs. 4--7, we show the dependences of the differential branching ratios on
the variation of (a) $C_X = \tilde{C}_L^V / C_L^V ({\rm SM})$, (b) $C_R^V / C_L^V ({\rm SM})$,
(c) $C_L^S / C_L^V ({\rm SM})$ and (d) $C_L^T / C_L^V ({\rm SM})$
for decays of $B \rightarrow K$ (Fig. 4),  $B \rightarrow K^*_{h=0}$ (Fig. 5),
$(B \rightarrow K^*_{h=+1}) + (B \rightarrow K^*_{h=-1}) $ (Fig. 6) and 
$B \rightarrow K^* $ (Fig. 7).
The thick solid line always indicates the SM case.
In Fig. 4, the dependence of the differential branching ratios as a function of
the missing mass-squared is shown for $B \rightarrow K$ decay. 
In  Fig. 4(a), the dashed, dotted and dot-dashed line
correspond to $\tilde{C}_L^V /C_L^V ({\rm SM}) = -0.7, 0.7, 1.0$ cases,
respectively.
In Fig. 4(b), the dashed, dotted and dot-dashed line
correspond to $C_R^V /C_L^V ({\rm SM}) = -0.7, 0.7, 1.0$ cases,
respectively.
In Fig. 4(c), the dashed and dotted line
correspond to $C_L^S /C_L^V ({\rm SM}) = \pm 0.7, \pm 1.0$ cases,
respectively.
In Fig. 4(d), the dashed and dotted line
correspond to $C_L^T /C_L^V ({\rm SM}) = \pm 0.7, \pm 1.0$ cases,
respectively.
We can see that the vector-type interactions change
the spectrum near the large-recoil limit ($q^2 \rightarrow 0$),  while the
scalar- and tensor-type interactions increase the spectrum in the
center of the phase space and do not change the spectrum at the
large-recoil limit. 

In Fig. 5, the dependence of the differential branching ratios as a function of
the missing mass-squared is shown for $B \rightarrow K^*_{h=0}$ decays. 
In Fig. 5(a), the dashed, dotted and dot-dashed line
correspond to $\tilde{C}_L^V /C_L^V ({\rm SM}) = -0.7, 0.7, 1.0$ cases,
respectively.
In Fig. 5(b), the dashed, dotted and dot-dashed line
correspond to $C_R^V /C_L^V ({\rm SM}) = -1.0, -0.7, 0.7$ cases,
respectively.
In Fig. 5(c), the dashed and dotted line
correspond to $C_L^S /C_L^V ({\rm SM}) = \pm 0.7, \pm 1.0$ cases,
respectively.
In Fig. 5(d), the dashed and dotted line
correspond to $C_L^T /C_L^V ({\rm SM}) = \pm 0.7, \pm 1.0$ cases,
respectively.
In this case too, the vector-type interaction changes the
spectrum for the large-recoil limit. However, the sign of the
contribution of $C_R$ is different from that of the $ B \rightarrow K$ case.
(see Fig. 5(b).)
The scalar-type interaction enhances the spectrum at the center (see Fig. 5(c))
and the tensor-type interaction enhances the spectrum near the zero-recoil
of $K^*$ (see Fig. 5(d)).

In Fig. 6, the dependence of the differential branching ratios as a function of
the missing mass-squared
for $(B \rightarrow K^*_{h=+1}) + (B \rightarrow K^*_{h=-1}) $ decays is shown.
In Fig. 6(a), the dashed, dotted and dot-dashed line
correspond to $\tilde{C}_L^V /C_L^V ({\rm SM}) = -0.7, 0.7, 1.0$ cases,
respectively.
In  Fig. 6(b), the dashed, dotted, dot-dashed
and solid line
correspond to $C_R^V /C_L^V ({\rm SM}) = -1.0, -0.7, 0.7, 1.0$ cases,
respectively.
Fig. 6(c) shows that there is no dependence on the $C_L^S$.
In Fig. 6(d), the dashed and dotted line
correspond to $C_L^T /C_L^V ({\rm SM}) = \pm 0.7, \pm 1.0$ cases,
respectively.

In Fig. 7, the dependence of the differential branching ratios as a function of
the missing mass-squared for $B \rightarrow K^*$ decays is shown, 
{\it i.e.} the sum of Fig. 5 and Fig. 6.
In Fig. 7(a), the dashed, dotted and dot-dashed line
correspond to $\tilde{C}_L^V /C_L^V ({\rm SM}) = -0.7, 0.7, 1.0$ cases,
respectively.
In Fig. 7(b), the dashed, dotted, dot-dashed
and solid line
correspond to $C_R^V /C_L^V ({\rm SM}) = -1.0, -0.7, 0.7, 1.0$ cases,
respectively.
In Fig. 7(c), the dashed and dotted line
correspond to $C_L^S /C_L^V ({\rm SM}) = \pm 0.7, \pm 1.0$ cases,
respectively.
In Fig. 7(d), the dashed and dotted line
correspond to $C_L^T /C_L^V ({\rm SM}) = \pm 0.7, \pm 1.0$ cases,
respectively.
As one can see from the Figs. 4-7, the various new physics interactions
have their own characteristic nature for the missing mass-squared spectrum.
Therefore, these spectra can be used to discriminate the various new physics effects.


To summarize, we presented the possible new physics effects on $ B \rightarrow
K^{(*)} \nu \nu$ decays in a model-independent way. With dimension-$6$
four-fermi interactions, 
not only the $B \rightarrow K^{(*)} \nu {\bar \nu} $ decay but also
the total lepton-number-violating $ B \rightarrow K^{(*)} {\nu} {\nu} $ or
$ B \rightarrow K^{(*)} {\bar{\nu}} {\bar{\nu}} $
decay may occur.
Using the form factor of Ref. \cite{MELI}, we have shown how the branching
ratios and the missing mass-squared spectrum depend on the new interactions.
We can infer from the Figures that the branching ratios and the spectrum
are useful for discriminating the various new physics effects systematically. 

\acknowledgements

\noindent
We thank G. Cvetic for careful reading of the manuscript and his
valuable comments. The work of C.S.K. was supported 
in part by KRF Non-Directed-Research-Fund, Project No. 1997-001-D00111,
in part by the BSRI Program, Ministry of Education, Project No. 98-015-D00061,
in part by the KOSEF-DFG large collaboration project, 
Project No. 96-0702-01-01-2.
The work of Y.G.K. was supported by KOSEF Postdoctoral Program.
T.M. would like to thank APCTP where we started this work.
The work of T.M. was supported by a Grand-in Aid for Scientific
Research on Priority Areas (Physics of CP violation). \\

\begin{appendix}
\section{Statistical Factors for Theoretically Indistinguishable Neutrinos}

\noindent
We derive the relative statistical factors for the decay 
$B \rightarrow K {\bar \nu_i}{\bar \nu_j}({\nu_i}{\nu_j})$
for the case of the (theoretically) indistinguishable neutrinos 
$i=j$ as compared to the case of the (theoretically)
distinguishable neutrinos $i \neq j$ (and $\nu \bar \nu$).
First we define the neutrino field as,
\begin{eqnarray}
\psi_L=\sum_p [a_p u_{Lp} \exp(-ipx) + b_p^\dagger u_{Lp} \exp(ipx)].
\label{A-1}
\end{eqnarray}
where $u_{Lp}$ is four-component spinor 
which has only the lowest two components ($\eta_p$) nonzero:
$u_{Lp}=(0, \eta_p)^T$.
By defining the final two anti-neutrino states as
$|{p1}^i ,{p2}^j >=b^{i\dagger}_{p1}b^{j\dagger}_{p2}|0>$, we obtain
\begin{eqnarray}
<p1^i,p2^j|{\overline{\psi_{Li}^C}}{\psi_{Lj}}|0> &=&
-i \delta_{ij}u_{Lj}(p2)^{t} \gamma_2 \gamma_0 u_{Li}(p1)+
+i u_{Li}(p1)^{t} \gamma_2 \gamma_0 u_{Lj}(p2) \\
&=& i(1+\delta_{ij})u_{Li}(p1)^{t} \gamma_2 \gamma_0 u_{Lj}(p2).
\label{A-2}
\end{eqnarray}
We can see that the matrix element for the indistinguishable neutrinos
is enhanced by a factor of two (and, therefore, a factor of four for the amplitude-squared) 
compared with  the matrix element for the distinguishable case.
After including a factor $1/2$ from the indistinguishable phase space, 
the decay rate for the indistinguishable neutrinos is twice larger than that of the
distinguishable neutrinos. 
Even though experimentally all the neutrinos are practically 
indistinguishable in those environments,
this factor two applies only to the theoretically indistinguishable case, 
{it i.e.},  $B \to K^{(*)} {\nu_i}{\nu_i}$ and $B \to K^{(*)} \bar{\nu_i}\bar{\nu_i}$.

\end{appendix}

\begin{table}
\caption{\label{t1}
Parameters of the fit 
$f_i (q^2)=f_i (0)/[1-\sigma_1 q^2+\sigma_2 q^4]$ to 
the $B \rightarrow (K, K^*)$ transition form factors 
in the GI-OGE model.\\}
\begin{tabular}{ccccccccccc}
Ref. & $f_+ (0)$ & $f_- (0)$ & $s(0)$ & 
$g(0)$ & $f(0)$ & $a_+ (0)$ & $a_- (0)$ & $h (0)$ & $g_+ (0)$ & $g_- (0)$ \\
& $\sigma_1 $ & $\sigma_1 $ & $\sigma_1 $ & $\sigma_1 $ & $\sigma_1 $ 
& $\sigma_1 $ & $\sigma_1 $ & $\sigma_1 $ & $\sigma_1 $ & $\sigma_1 $ \\
& $\sigma_2 $ & $\sigma_2 $ & $\sigma_2 $ & $\sigma_2 $ & $\sigma_2 $ 
& $\sigma_2 $ & $\sigma_2 $ & $\sigma_2 $ & $\sigma_2 $ & $\sigma_2 $ \\
\hline
GI-OGE & 0.33 & -0.27 & 0.057 & 0.063 & 2.01 & -0.0454 & 0.053 & 0.0056 & -0.3540 & 0.313 \\
& 0.0519 & 0.0524 & 0.0517 & 0.0523 & 0.0212 & 0.039 & 0.044 & 0.0657 & 0.0523 & 0.053 \\
& 0.00065 & 0.00066 & 0.00064 & 0.00066 & 0.00009 & 0.00004 & 0.00023 & 0.0010 & 0.0007 & 0.00067 \\
\end{tabular}
\end{table}

\begin{figure}[ht]
\hspace*{-1.0 truein}
\psfig{figure=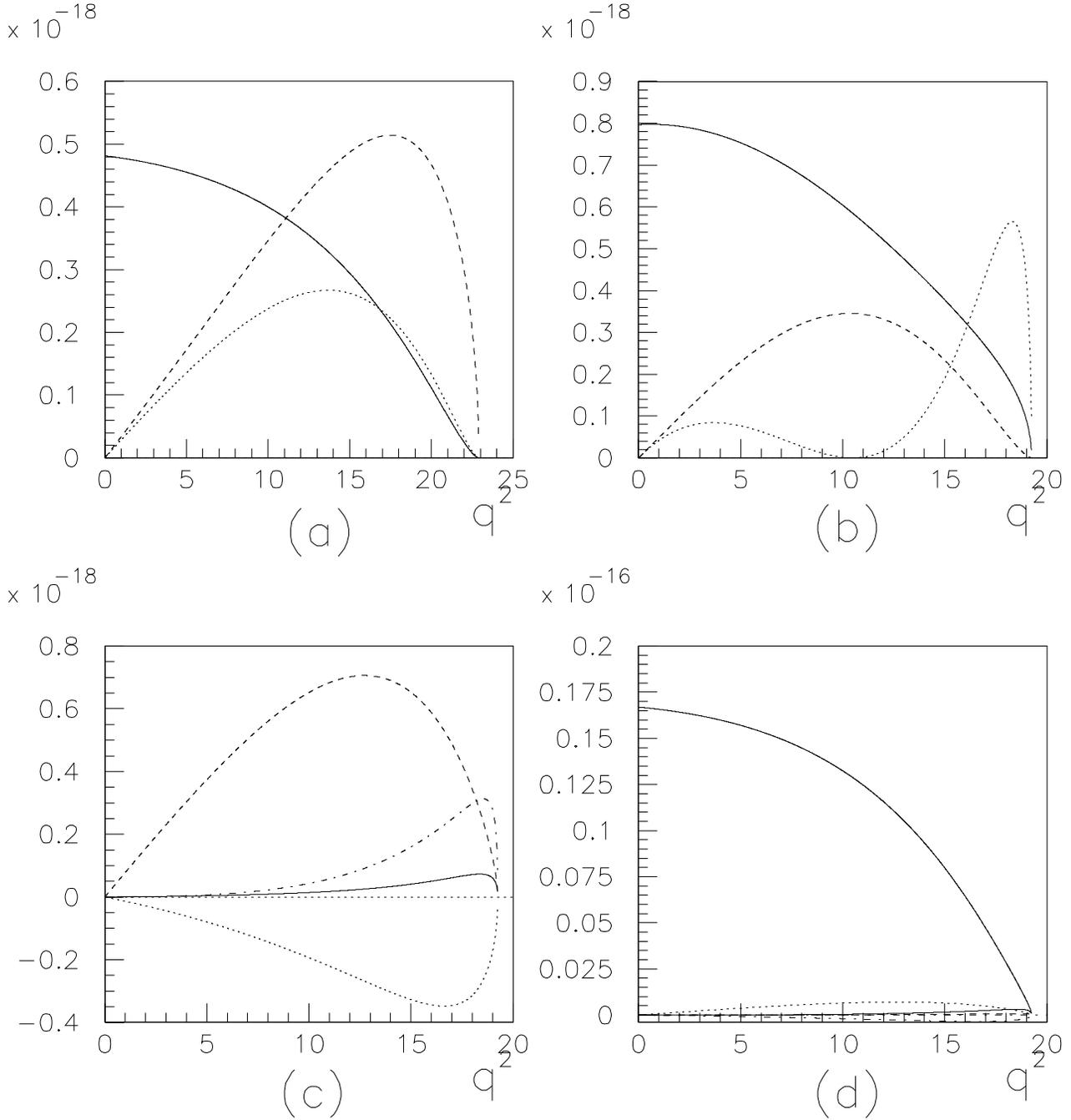}
\caption{$q^2$ (in GeV$^2$ scale) dependence of various factorized
functions (in GeV$^{-1}$ scale) for (a) $(B \rightarrow K)$, (b) $(B \rightarrow K^*)_L$,
(c) and (d) $(B \rightarrow K^*)_T$. 
The functions are defined in Section 3. 
See text for the details.}
\label{q2all}
\end{figure}

\begin{figure}[ht]
\hspace*{-1.0 truein}
\psfig{figure=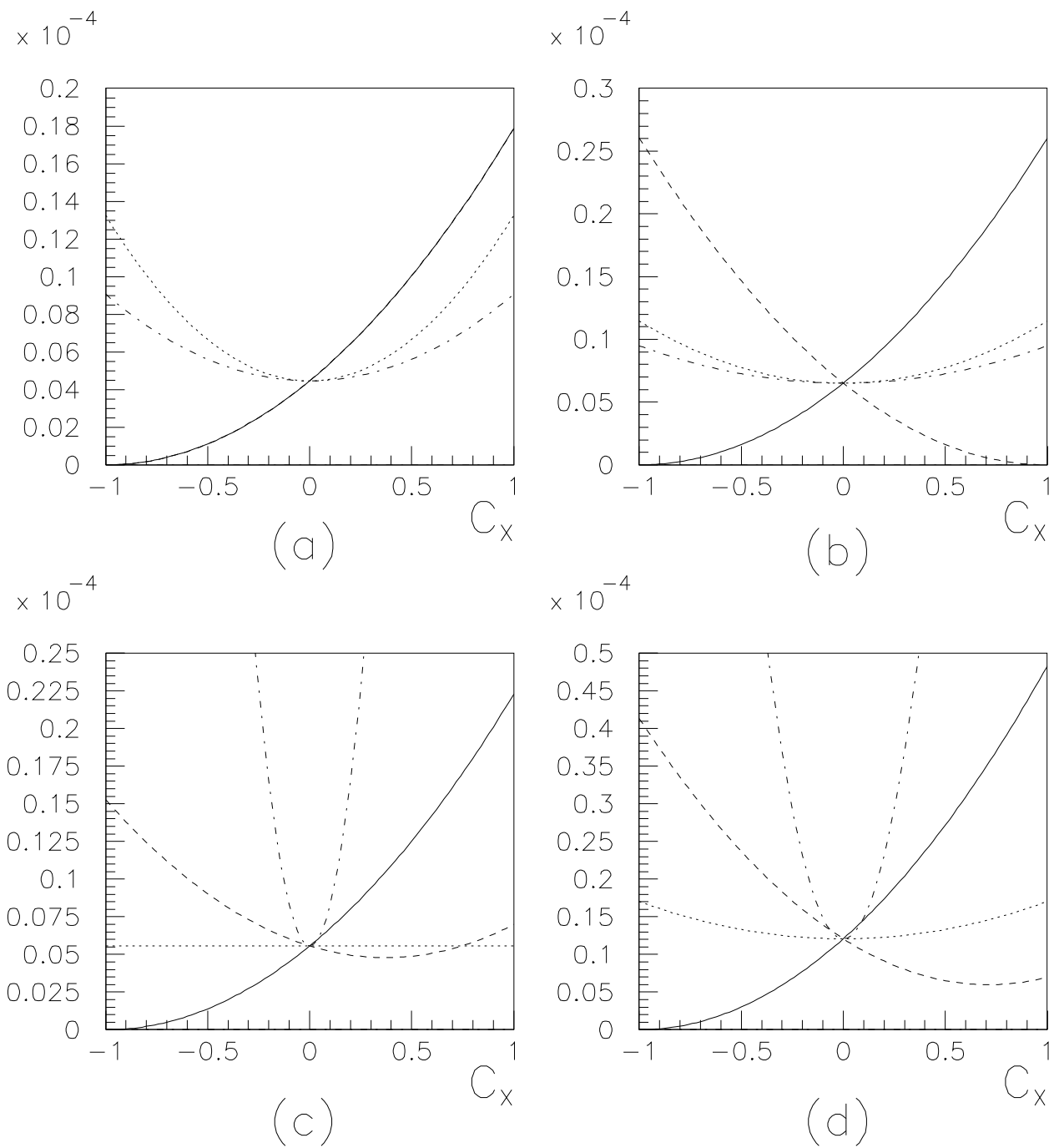}
\caption{The dependence of the branching ratios (a) ${\cal BR}(B \rightarrow K)$,
(b) ${\cal BR}(B \rightarrow K^*)_L$, 
(c) ${\cal BR}(B \rightarrow K^*)_T$ 
and (d) ${\cal BR}(B \rightarrow K^*)$ 
on the coefficients $C_X$. 
The dependence on $C_X = \tilde{C}_L^V / C_L^V ({\rm SM})$, $C_R^V / C_L^V ({\rm SM})$,
$C_L^S / C_L^V ({\rm SM})$
and $C_L^T / C_L^V ({\rm SM})$ corresponds to 
the solid, dashed, dotted
and dot-dashed line, respectively, where
$\tilde{C}_L^V \equiv C_L^V - C_L^V ({\rm SM})$.
}
\label{brall}
\end{figure}

\begin{figure}[ht]
\hspace*{-1.0 truein}
\psfig{figure=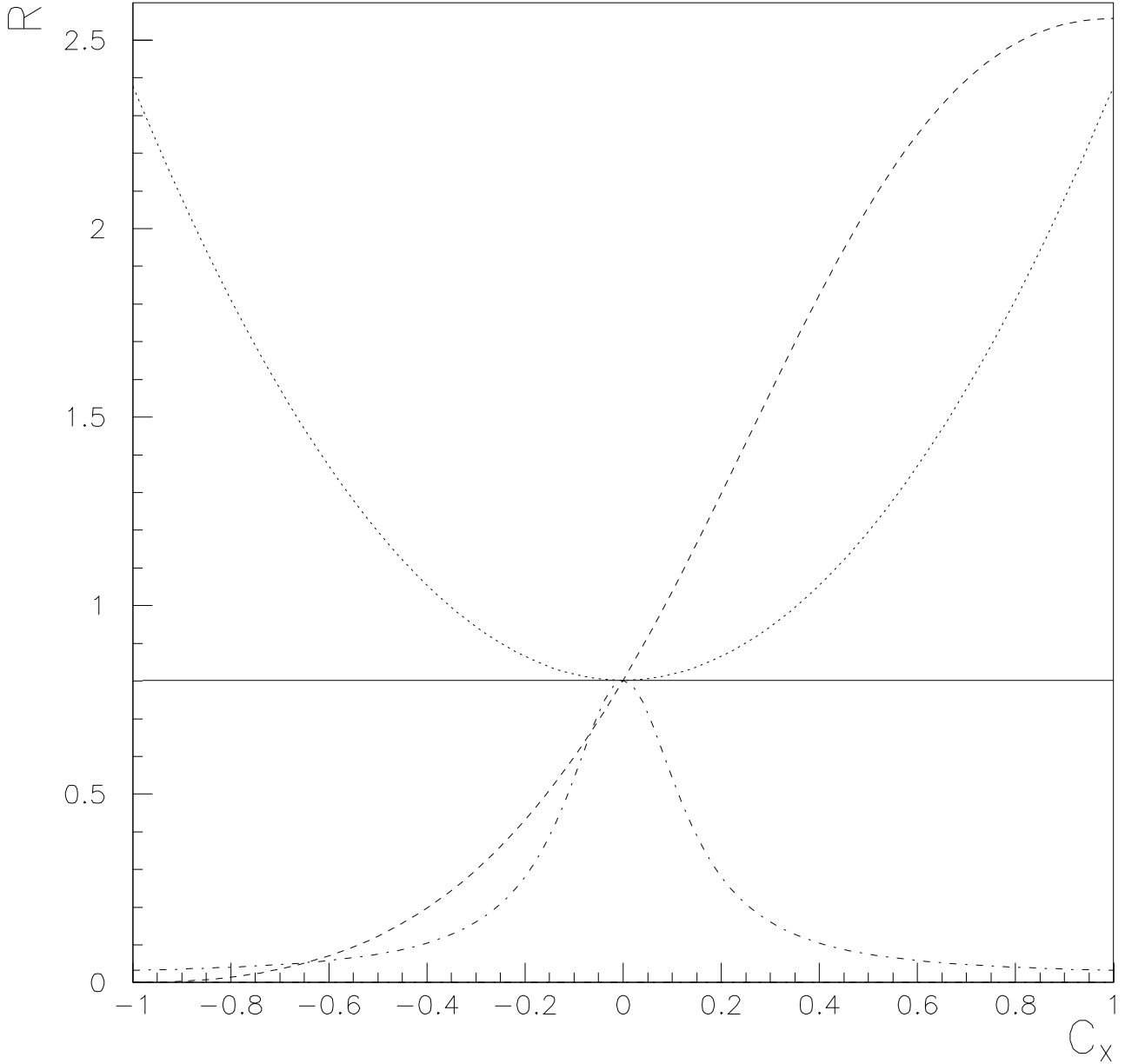}
\caption{The dependence of the ratio $R$ (defined in Eq. (38)) on $C_X$.
The dependence on $C_X = \tilde{C}_L^V / C_L^V ({\rm SM})$, $C_R^V / C_L^V ({\rm SM})$,
$C_L^S / C_L^V ({\rm SM})$
and $C_L^T / C_L^V ({\rm SM})$ corresponds to 
the solid, dashed, dotted
and dot-dashed line, respectively.}
\label{atr}
\end{figure}

\begin{figure}[ht]
\hspace*{-1.0 truein}
\psfig{figure=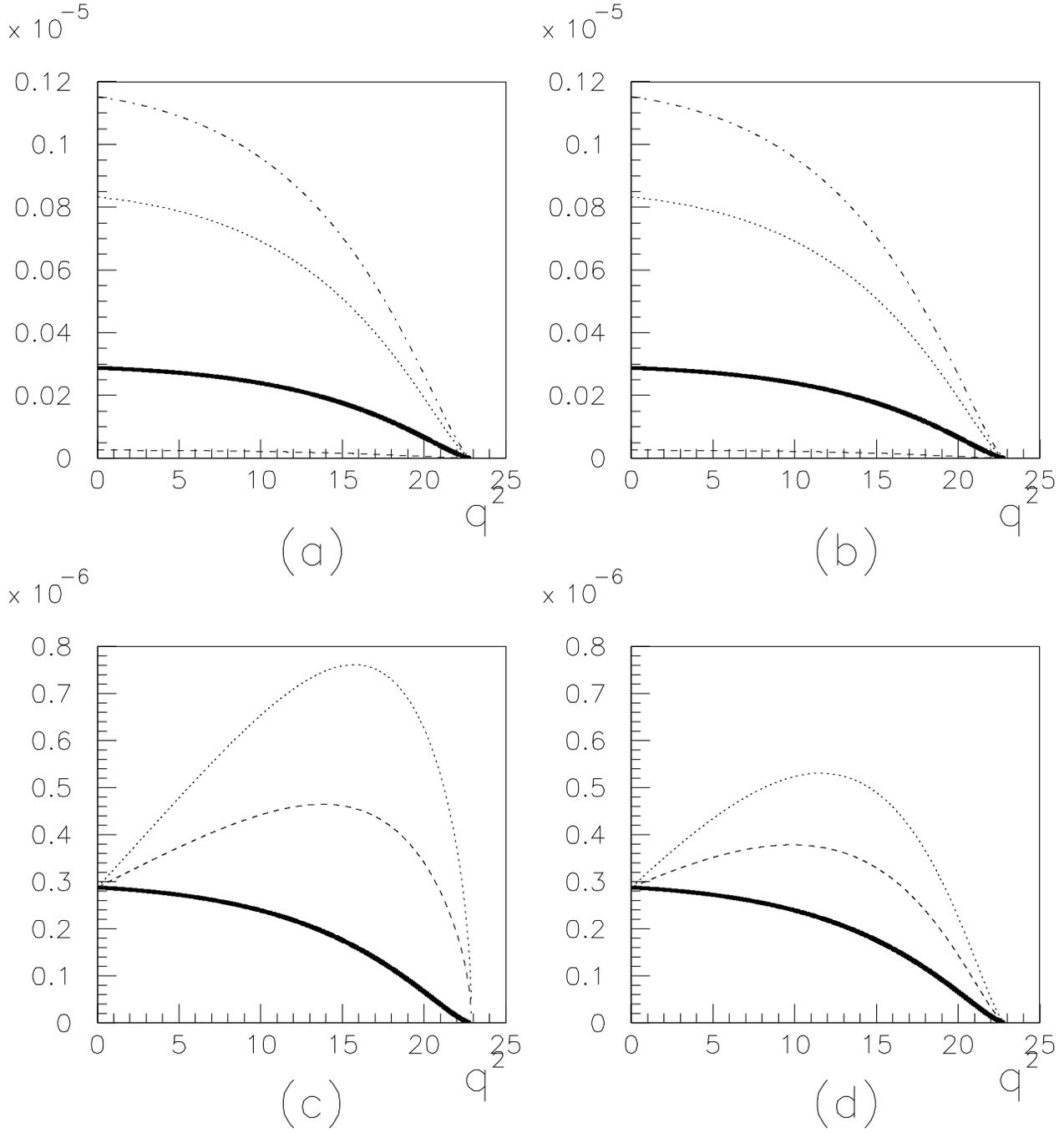}
\caption{The differential branching ratios for $(B \rightarrow K)$
depending on (a) $C_X=\tilde{C}_L^V / C_L^V ({\rm SM})$, (b) $C_R^V / C_L^V ({\rm SM})$,
(c) $C_L^S / C_L^V ({\rm SM})$
and (d) $C_L^T / C_L^V ({\rm SM})$. The thick solid line indicates the SM case.
See text for the numerical variation of $C_X$.}
\label{bk}
\end{figure}

\begin{figure}[ht]
\hspace*{-1.0 truein}
\psfig{figure=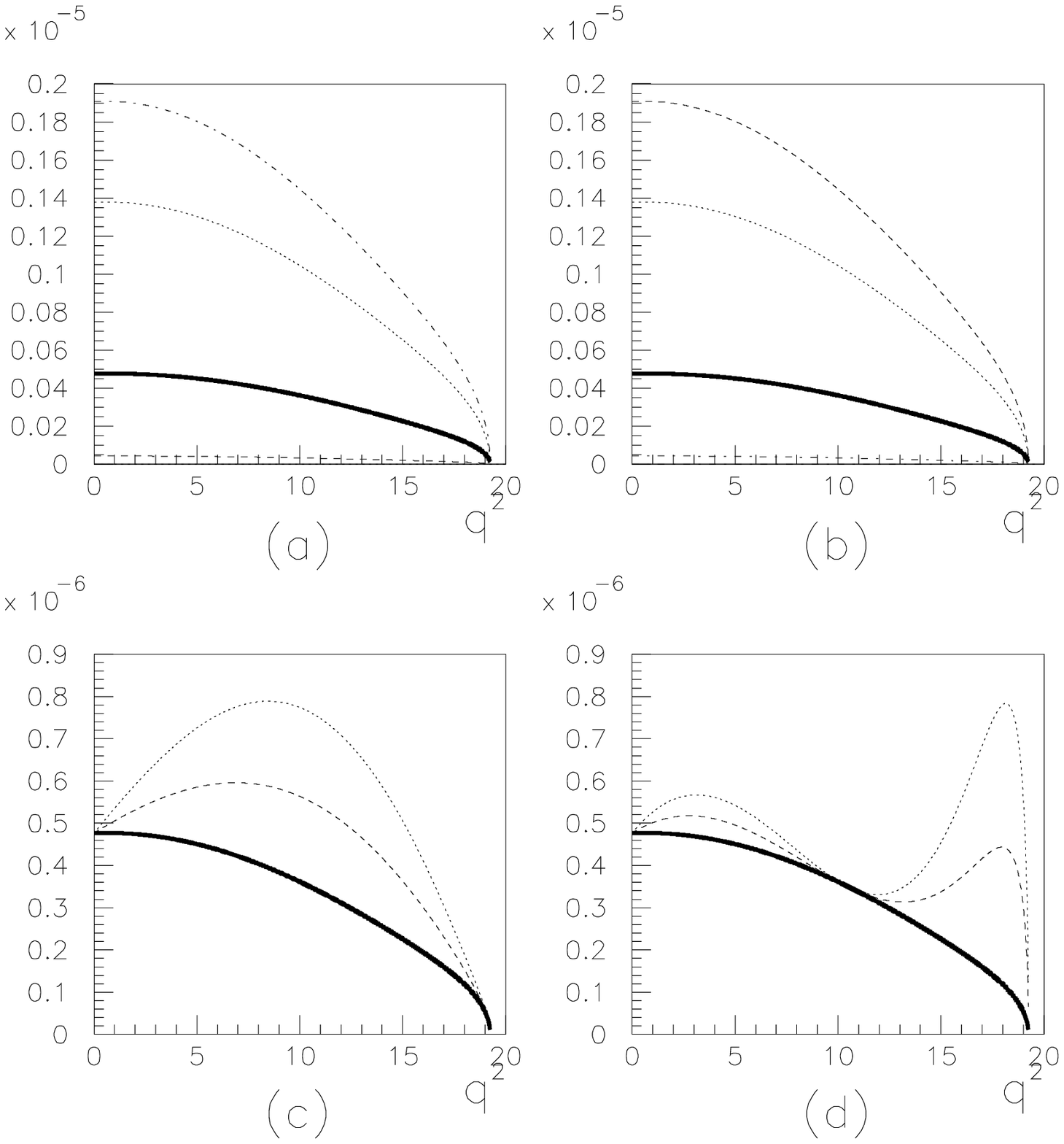}
\caption{The differential branching ratios 
for $(B \rightarrow K^*)_L$ as in Fig. 4.}
\label{bkl}
\end{figure}

\begin{figure}[ht]
\hspace*{-1.0 truein}
\psfig{figure=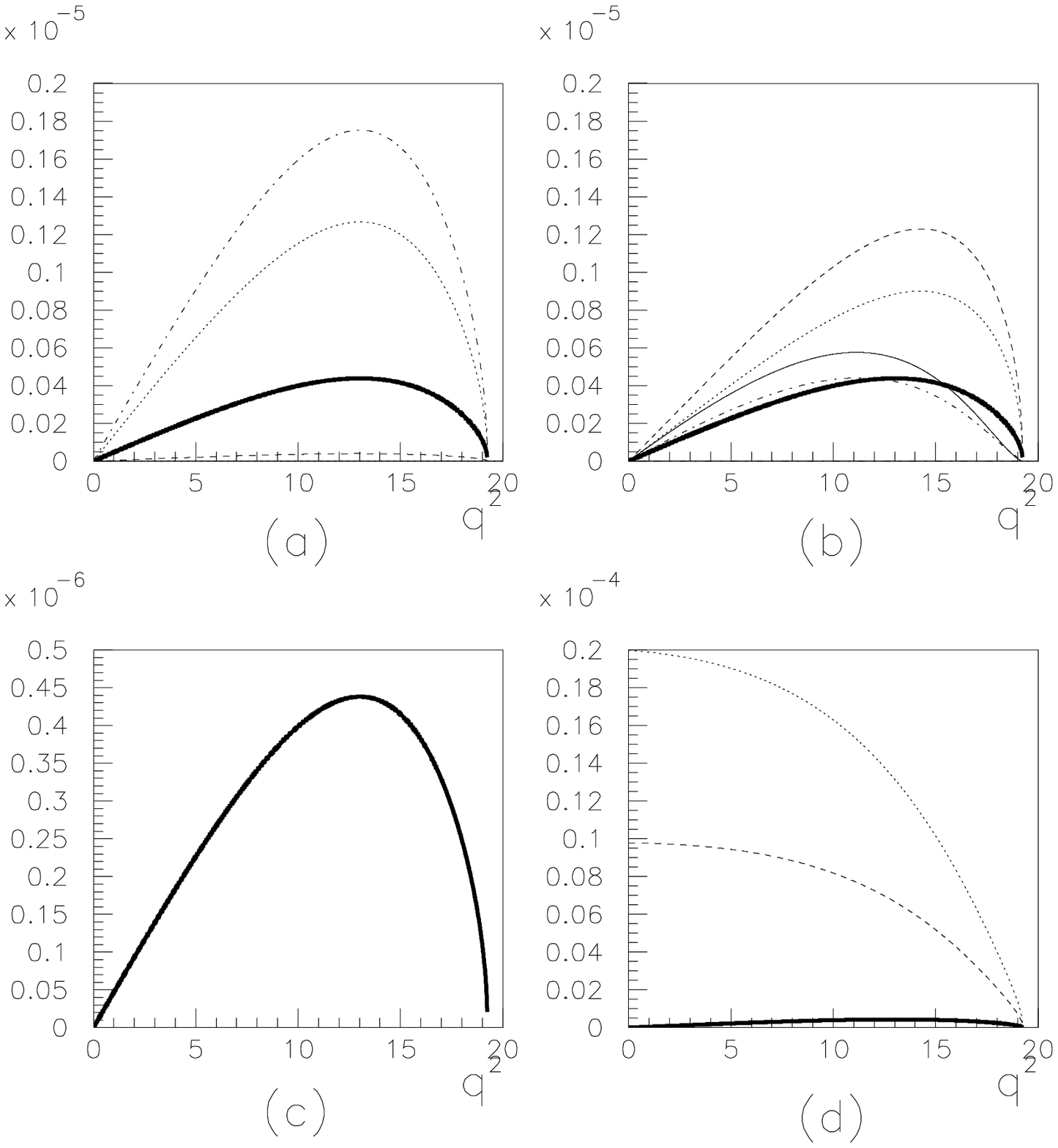}
\caption{The differential branching ratios 
for $(B \rightarrow K^*)_T$ as in Fig. 4.}
\label{bkp}
\end{figure}

\begin{figure}[ht]
\hspace*{-1.0 truein}
\psfig{figure=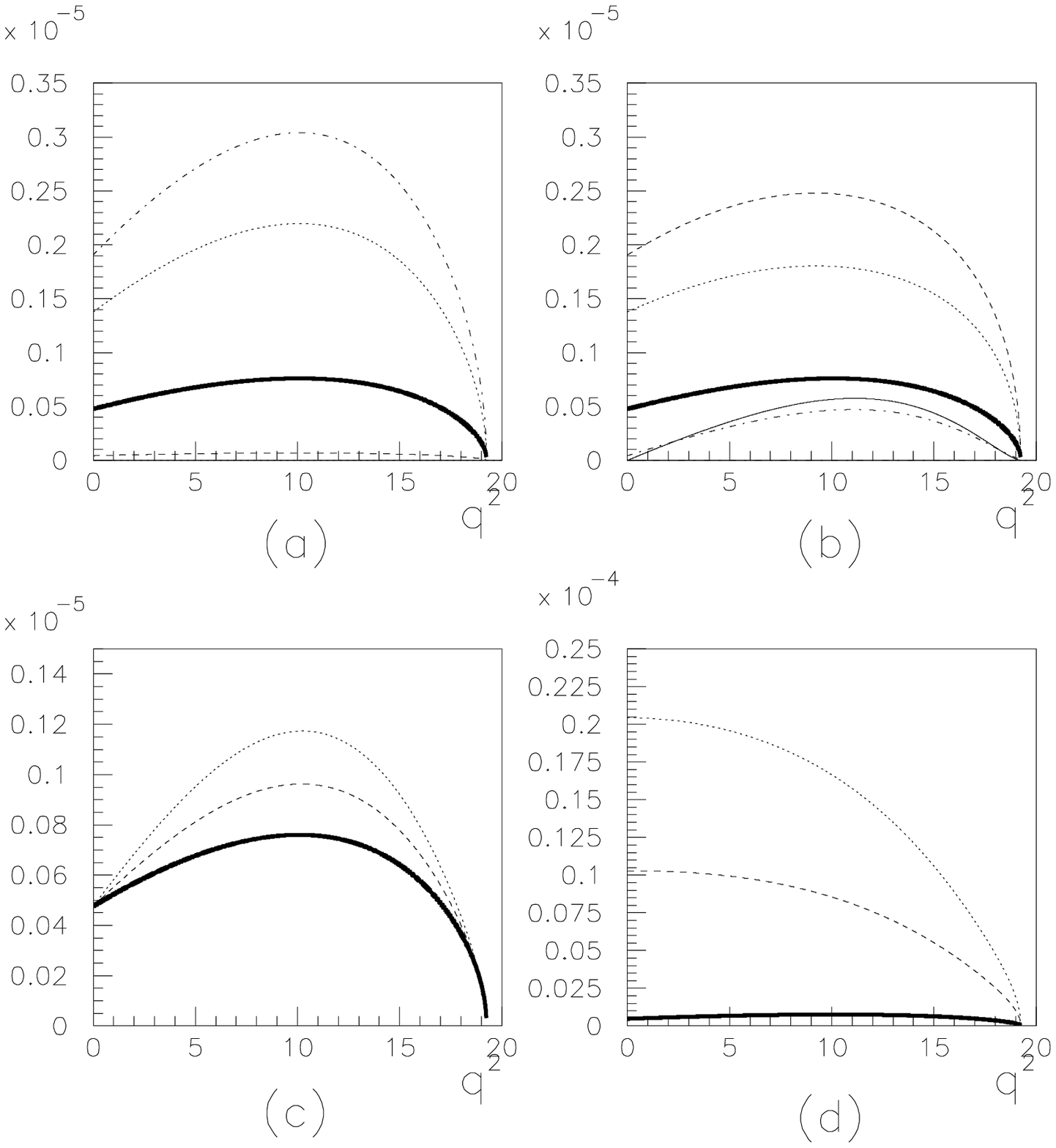}
\caption{The differential branching ratios 
for $(B \rightarrow K^*)$ as in Fig. 4.}
\label{bkm}
\end{figure}


\begin{references}
\bibitem{buras}
 G. Buchalla and A. Buras,
 {\it Nucl. Phys.} {\bf B400} (1993) 225; {\it Phys. Rev.} {\bf D54} (1996) 6782.
\bibitem{nir}
 Y. Grossman and Y. Nir, {\it Phys. Lett.} {\bf 398} (1997) 163.
\bibitem{grossman} 
 Y. Grossman, Z. Ligeti and E. Nardi,
 {\it Nucl. Phys.} {\bf B465} (1996) 369;  Erratum {\it ibid} {\bf B480} (1996) 753.
\bibitem{aliev} 
 T. M. Aliev and C. S. Kim,
 {\it Phys. Rev.} {\bf D58} (1998) 013003.
\bibitem{meli1}
 D. Melikov, N. Nikitin and S. Simula,
 {\it Phys. Lett.} {\bf B428} (1998) 171.
\bibitem{CKIM}
 C. S. Kim, S. Fukae, T. Morozumi and T. Yoshikawa,
 {\it Phys. Rev.} {\bf D59} (1999) 074013.
\bibitem{KIYO}
 Y. Kiyo, T. Morozumi, P. Parada, M.N. Rebelo and M. Tanimoto,
 {\it Prog. Theor. Phys} {\bf 101} (1999) 671.
\bibitem{MELI}
 D. Melikov, N. Nikitin and S. Simula,
 {\it Phys. Rev.} {\bf D57} (1998) 6814.
\bibitem{GI}
 S. Godfrey and N. Isgur,
 {\it Phys. Rev.} {\bf D32} (1985) 189.
\bibitem{PDBook} 
 Particle Data Group, Eur. Phys. J. {\bf C3} (1998) 1.

\end{references}
\end{document}